\documentclass[letterpaper]{article} 
\usepackage[]{aaai23}  
\usepackage{times}  
\usepackage{helvet}  
\usepackage{courier}  
\usepackage[hyphens]{url}  
\usepackage{graphicx} 
\usepackage{natbib}  
\usepackage{caption} 
\usepackage{algorithm}
\usepackage{algorithmic}

\usepackage{newfloat}
\usepackage{listings}
\usepackage{soul}

\usepackage[table,xcdraw]{xcolor}

\DeclareCaptionStyle{ruled}{labelfont=normalfont,labelsep=colon,strut=off} 
\floatstyle{ruled}
\newfloat{listing}{tb}{lst}{}
\floatname{listing}{Listing}
\usepackage{tablefootnote}
\title{Understanding Ethics, Privacy, and Regulations in Smart Video Surveillance for Public Safety}

\author {
    Babak Rahimi Ardabili\textsuperscript{\rm 1},
    Armin Danesh Pazho\textsuperscript{\rm 1},
    Ghazal Alinezhad Noghre\textsuperscript{\rm 1},
    Christopher Neff\textsuperscript{\rm 1},
    Arun Ravindran\textsuperscript{\rm 1},
    Hamed Tabkhi\textsuperscript{\rm 1}   
}
\affiliations {
    \textsuperscript{\rm 1} brahimia@uncc.edu, adaneshp@uncc.edu, galinezh@uncc.edu, cneff1@uncc.edu, arun.ravindran@uncc.edu, htabkhiv@uncc.edu \\ University of North Carolina at Charlotte, Charlotte NC 28223, USA
}

\begin{document}

\maketitle

\begin{abstract}
Recently, Smart Video Surveillance (SVS) systems have been receiving more attention among scholars and developers as a substitute for the current passive surveillance systems. These systems are used to make the policing and monitoring systems more efficient and improve public safety. However, the nature of these systems in monitoring the public's daily activities brings different ethical challenges. There are different approaches for addressing privacy issues in implementing the SVS. In this paper, we are focusing on the role of design considering ethical and privacy challenges in SVS. Reviewing four policy protection regulations that generate an overview of best practices for privacy protection, we argue that ethical and privacy concerns could be addressed through four lenses: algorithm, system, model, and data. As an case study, we describe our proposed system and illustrate how our system can create a baseline for designing a privacy perseverance system to deliver safety to society. We used several Artificial Intelligence algorithms, such as object detection, single and multi camera re-identification, action recognition, and anomaly detection, to provide a basic functional system. We also use cloud-native services to implement a smartphone application in order to deliver the outputs to the end users.
\end{abstract}

\section{Introduction}
The recent improvements in Artificial Intelligence (AI) and Machine Learning (ML) algorithms dramatically affected different areas of science, society, and daily life. The effects are not limited to the novel approaches to address the classic problems but broaden the scope of the issues that can be addressed in each domain. One of the areas that AI has impacted is Smart Video Surveillance (SVS). With the recent developments in Computer Vision algorithms, faster processors, ubiquitous broadband, and inexpensive cameras, the surveillance process has become more intelligent and efficient\cite{atitallah2020leveraging}. Such a capacity has led citizens, municipalities, and researchers to use SVS in various applications. The SVS systems can help law enforcement to enhance society's safety as a desirable public interest by improving situational awareness\cite{zhang2014context}. Some major applications of SVS are anomaly detection, criminal behavior identification, pedestrian safety, and public safety. As a result of the potential of the technology, the global video surveillance market is expected to have a Compound Annual Growth Rate (CAGR) of 14.6 percent from 2020 to 2027, which means the 42.94 billion USD value in 2019 to reach 144.85 billion USD by 2027 \cite{fathy2022integrating}.  

The rapid development of SVS has resulted in some ethical debates. The ethical challenges in this context can be understood through two approaches: approaches toward acceptance of the technology and approaches toward implementing the technologies\cite{hartzog2018privacyos}. Investigating the acceptance of technology brings ethical issues such as trust. To adopt a technology, most of the potential users, including public entities, private entities, and citizens, need to trust the technology\cite{schomakers2021users}. On the other hand, implementing a technology raises additional ethical challenges such as data privacy and data security\cite{abuhammad2020covid}. Emerging technology-based data-driven economy deepened the trade-off between individual privacy rights and the social goals of the technology\cite{waldman2022privacy}. Privacy concerns can be addressed from different perspectives. However, addressing this issue at the designing level provides a more flexible and efficient solution toward the issue\cite{hartzog2018privacyos}. The main goal of implementing SVS systems is to achieve a reasonable level of safety in society. According to \cite{nissenbaum2004privacy}, privacy concerns are among the most important yet less approached ethical challenges in the video surveillance sector. This controversial challenge of the topic impedes the technology from being used vastly in the community, despite the significant advances in Computer Vision and its promising potential.

Policymakers regulate technologies to assure the public that their ethical concerns have been addressed in technological advancement\cite{leenes2019regulating}. A technology regulation not only requires to be dynamic but also should be domain specific \cite{boekaerts2000self} to effectively provide a solution for society's concerns. The controversial nature of the SVS field from the privacy perspective, besides the lack of federal regulations and domain specifics nature of the privacy policy in the US \cite{acquisti2015privacy, nissenbaum2004privacy}, brings the requirement of a particular focus on SVS policies and regulations. State-by-state approach toward privacy protection policies resulted in banning facial recognition technology, one of the leading technologies in the domain of SVS, in some states. For example, Berkeley and San Francisco in California state banned the use of facial recognition. Banning the use of facial recognition technology is not limited to California. Since 2020 Vermont, Virginia, Massachusetts, Maine, New York, Washington, Maryland, and Oregon states banned or limited the use of facial recognition technology in law enforcement. These regulations and policies affected big technology companies such as IBM, Microsoft, and Facebook selling and using facial recognition technologies and tools \cite{almeida2022ethics}.

In this paper, we propose a SVS system design. The system is designed to use the videos from existing cameras in public places to improve public safety. Although we are not the only research team using this setup for this purpose, we argue that the different elements of the system are selected intentionally to provide a baseline solution to the ethical challenge of privacy. Our approach toward this challenge shows the importance of design in addressing social issues such as privacy. 

\section{Related Work}
\subsection{Smart Video Surveillance in Research}
SVS has been a hot research topic for many years. In 2015, researchers at the University of Alcal\'a \cite{MallBehaviors} proposed a system for real-time detection of suspicious behaviors in shopping malls, using a combination of artificial intelligence approaches to track individuals through a mall's security system and determine when suspicious behaviors occur. However, like many works in the field, no regard is paid to ethical concerns, the privacy of the people being tracked, or any bias that might be learned by the system. Instead, the research focuses on real-time execution and achieving high accuracy on publicly available datasets. More recently, Peeking into the Future \cite{Peeking} proposed an end-to-end system for predicting the actions of people in a video surveillance setting. While the work itself does not pay any mind to ethical, privacy, or fairness concerns, they conclude that future work geared towards real-world applications may need to consider these issues as a priority.

In REVAMP\textsuperscript{2}T \cite{REVAMP2T}, a focus is put on performing person re-identification and tracking in a multi-camera environment while preserving the privacy of the persons being tracked. They propose two policies to achieve this. The first is that no image data is stored or transferred across the network; the system destroys the image data as soon as it is processed. They argue that this prevents even individuals with direct access to the system from accessing an individual's personally identifiable information. The second is that instead of using invasive technologies that identify individuals (e.g. facial recognition) their re-identification algorithm uses an abstract representation of a person's features that is uninterpretable by humans. In this way, they aim to focus on differentation between people instead of personal identification. Other works have taken similar approaches and applied them directly to the field of SVS \cite{TX2MTMC}.

\subsection{Smart Video Surveillance in Industry}
SVS systems have been used vastly in different sectors. Although some industrial use cases exist for these systems, addressing privacy issues is not the primary goal of their proposed systems. Most of the firms in the SVS sector offer different video management services. In most cases, they offer a built-in feature integrated into their general service as their security solution.    

Some firms offer services to blur the actual videos. Milestone systems, for example, offer a "privacy masking" feature to protect privacy. This feature provides a modular blurring algorithm. The user can opt out of blurring the video and choose the intensity of the mask\footnote{https://www.milestonesys.com/}. In this setup, the actual videos are still accessible. 

A couple of enterprises provide SVS-based solutions to detect crime. They offer object and person detection and action recognition services. They rarely provide information on the type of data and algorithms they are using\footnote{https://getsafeandsound.com/2021/01/top-video-surveillance-companies-2021/}. Avigilion corporation is an example of these corporations. Avigilion provides a search feature that enables the clients to search images for specific individuals and license plates \footnote{https://www.avigilon.com/}. Misuse of this feature might violate the privacy rights of the individual.

Some firms offer privacy perseverance systems but need to provide clear information on their approach toward designing a privacy perseverance system. Genetec Omnicast, for instance, delivers video surveillance management services to clients. Genetec mentions privacy and security protection as the feature of the service \footnote{https://www.genetec.com/}. However, they mainly discussed the cyber-security features of the products.

\begin{table*}[]
\centering
\caption{Summary of privacy protection acts (GDPR\cite{viorescu20172018}, HIPAA\cite{hipaa}, CCPA\cite{10.2307/j.ctvjghvnn}, ADPPA is derived from https://www.congress.gov/.}
\label{tab:sum}
\resizebox{\textwidth}{!}{%
\begin{tabular}{c|c|c|c|c|c|c}
\textbf{Regulations} & \textbf{Domain}                                                & \textbf{Type of Data}                                     & \textbf{Data Collection}                                              & \textbf{Data Transfer}                                                           & \textbf{Data Processing}                                              & \textbf{Data Retention}                                                           \\ \hline \hline
GDPR                 & General                                                        & Personal data                                             & Minimized                                                             & \begin{tabular}[c]{@{}c@{}}By user's \\ contest\end{tabular}                     & \begin{tabular}[c]{@{}c@{}}Consistent with\\ the purpose\end{tabular} & \begin{tabular}[c]{@{}c@{}}Consistent with\\ the purpose\end{tabular}             \\ \hline
HIPAA                & \begin{tabular}[c]{@{}c@{}}Health \& \\ Insurance\end{tabular} & \begin{tabular}[c]{@{}c@{}}Medical\\ records\end{tabular} & \begin{tabular}[c]{@{}c@{}}Consistent with\\ the purpose\end{tabular} & \begin{tabular}[c]{@{}c@{}}Allowed \\ between authorized\\ entities\end{tabular} & \begin{tabular}[c]{@{}c@{}}Allowed\\ by covered entities\end{tabular} & \begin{tabular}[c]{@{}c@{}}No fewer \\ than six years\end{tabular}                \\ \hline
CCPA                 & \begin{tabular}[c]{@{}c@{}}General\\ CL residents\end{tabular} & Personal data                                             & \begin{tabular}[c]{@{}c@{}}By informing\\ the consumers'\end{tabular} & \begin{tabular}[c]{@{}c@{}}Allowed by\\ prior notice\end{tabular}                & \begin{tabular}[c]{@{}c@{}}By \\ pseudonymization\end{tabular}        & \begin{tabular}[c]{@{}c@{}}By consumer's\\ request\end{tabular}                   \\ \hline
\rowcolor[HTML]{C0C0C0} 
ADPPA  & General    & Personal data       & Minimized                                                             & \begin{tabular}[c]{@{}c@{}}Deidentified\\ data are \\ allowed\end{tabular}       & \begin{tabular}[c]{@{}c@{}}Consistent with\\ the purpose\end{tabular} & \begin{tabular}[c]{@{}c@{}}At the end \\ of the service \\ or by law\end{tabular} \\ 
\end{tabular}
}
\end{table*}

\section{Privacy Perseverance System Features}
It is important to note that there has yet to be any federal law that addresses privacy issues from a technical perspective. However, some regulations have been developed to help developers ensure that the technology complies with public privacy concerns. In the US, the Health Insurance Portability and Accountability Act (HIPAA), the California Consumer Privacy Act (CPPA), and the American Data Privacy and Protection Act (ADPPA) are the most important acts that address this issue in different sectors. On the other hand, the General Data Protection Regulation (GDPR), the European Union's set of data privacy and protection rules, is the most noticeable act in Europe. 

The HIPAA Privacy Rule is a national standard that protects individuals' medical records and other identifiable health information. HIPAA applies to health plans, healthcare provider centers, and all providers utilize electronic healthcare transactions. The focus of HIPAA is to set specific rules to protect the privacy of individuals. Generally, HIPAA sets standards to protect individuals' rights to ban any use of health information without their authorization and to have access to their medical and health records to have a copy, to request corrections, and to transmit the electronic version to a third party. \cite{hipaa}.
California passed the first comprehensive commercial privacy law in 2018, the California Consumer Privacy Act (CCPA). The CPPA will provide clear guidelines for organizations and consumers in California and is applied to any for-profit entity doing business in California that collects, shares, or sells California consumers' personal data and has annual gross revenues of more than 25 million USD. The CCPA does not cover Protected Health Information (PHI) collected by covered entities or business associates and leaves it as HIPAA's subject. It also exempts medical information subject to California's analogous law, the Confidentiality of Medical Information Act (CMIA). In 2020, the California Consumer Privacy Rights Act (CPRA) was passed. CPRA Expands CPPA and lets consumers to 
(1) ask businesses not to share personal information; (2) ask to modify their incorrect personal information; and (3) limit enterprises' usage of "sensitive personal information," including geolocation; race; ethnicity; religion; genetic data; private communications; sexual orientation; and specified health information. 

The American Data Privacy and Protection Act (ADPPA) has yet to be passed by congress; however, it is expected to be effective shortly. Therefore, investigating the ADPPA perspective toward privacy as the latest act is critical. The bill would apply to most entities, including nonprofits, common carriers, large data holders, and service providers. The ADPPA would regulate how organizations keep and use consumer data. According to this act, data collectors must minimize the data they collect unless it is "necessary, proportionate, and limited to" their business purpose. 
ADPPA specifically applies limitations on the transfer and, in some cases, processing of Social Security numbers, precise geolocation, biometric and genetic data, passwords, browsing history, and physical activity tracking\footnote{All related content derived from https://www.congress.gov/}.  

The General Data Protection Regulation (GDPR) is a data and privacy protection regulation of the European Union (EU)  in the EU and the European Economic Area (EEA). The GDPR aims to set standards to ensure individuals control their personal data and simplify the international business regulatory environment. The GDPR's regulations cover all "personal data." According to GDPR, personal data "includes any information relates to a living, identified, or identifiable person". Name, SSN, other identification numbers, location data, IP addresses, online cookies, images, email addresses, and content generated by the data subject are examples of personal data \cite{viorescu20172018}.
As we can see, these regulations are outside the SVS context. However, they provide a reasonable starting point to evaluate and address the privacy issues in the SVS context.

None of the discussed regulations are specifically designed to consider the privacy issue in the SVS context. However, summarizing the existing regulations and policy perspectives provide a framework to address the privacy concerns in the context. According to the previous section, privacy could be addressed from the algorithm, system, model, and data perspectives. 

As it is shown in Table \ref{tab:sum}, all acts banned covering entities from using identifiable information. Therefore, from the algorithm perspective, the best algorithms do not depend on identifiable information. The system should be designed to ensure that information is not transferred to a third party. This system complies with all the acts. These acts also mention the data retention and irreversibility of the data. The models that are used should address these two concerns. Finally, the type of data is essential. They should be de-identified data. Therefore, in a compliance system design, personally identifiable information (PII) and facial recognition technology should not be used.

\begin{figure*}[h]
        \centering
               \includegraphics[width=1\linewidth]{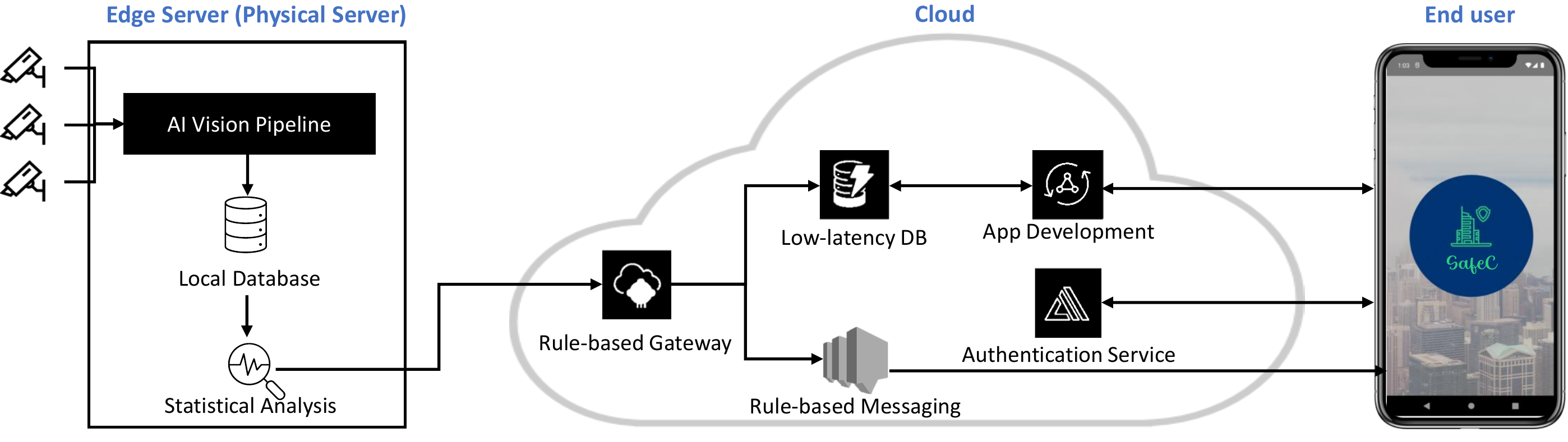}
                \caption{End-to-end system design.}
                \label{fig:view_points}
\end{figure*}

\section{Proposed System}

This section proposes our very early design of the AI-based SVS system. The goal of designing this system is to help the community improve public safety. Since the pre-installed cameras have access to images of people tracked by the cameras, it is crucial to consider the privacy issues in different system elements. 

The object of the proposed system is to use the videos from existing cameras in public areas and, by using AI algorithms, extract and deliver the information to the end users through a smartphone application. To fulfill the goal of the system, which is improving public safety, the community will need information such as detected objects, detected actions, detected anomalous behavior, and statistical data of the population. This system also needs to be equipped with a notifying system in an emergency. Figure \ref{fig:view_points} shows the end-to-end system design and how the different sections of the system work together.  

As shown in Figure \ref{fig:view_points}, the system has two main parts: an edge/physical server and a cloud-based server. All AI algorithms, and statistical models are run on the edge/physical server. The results of analyzed data are pushed to the cloud server. The cloud service is used to host the smartphone application and ensure that we will avoid facing technical and executive problems by increasing the number of users. The end user uses his/her device to check the stats in their desired location. 

\begin{figure*}[]
        \centering
               \includegraphics[width=1\linewidth]{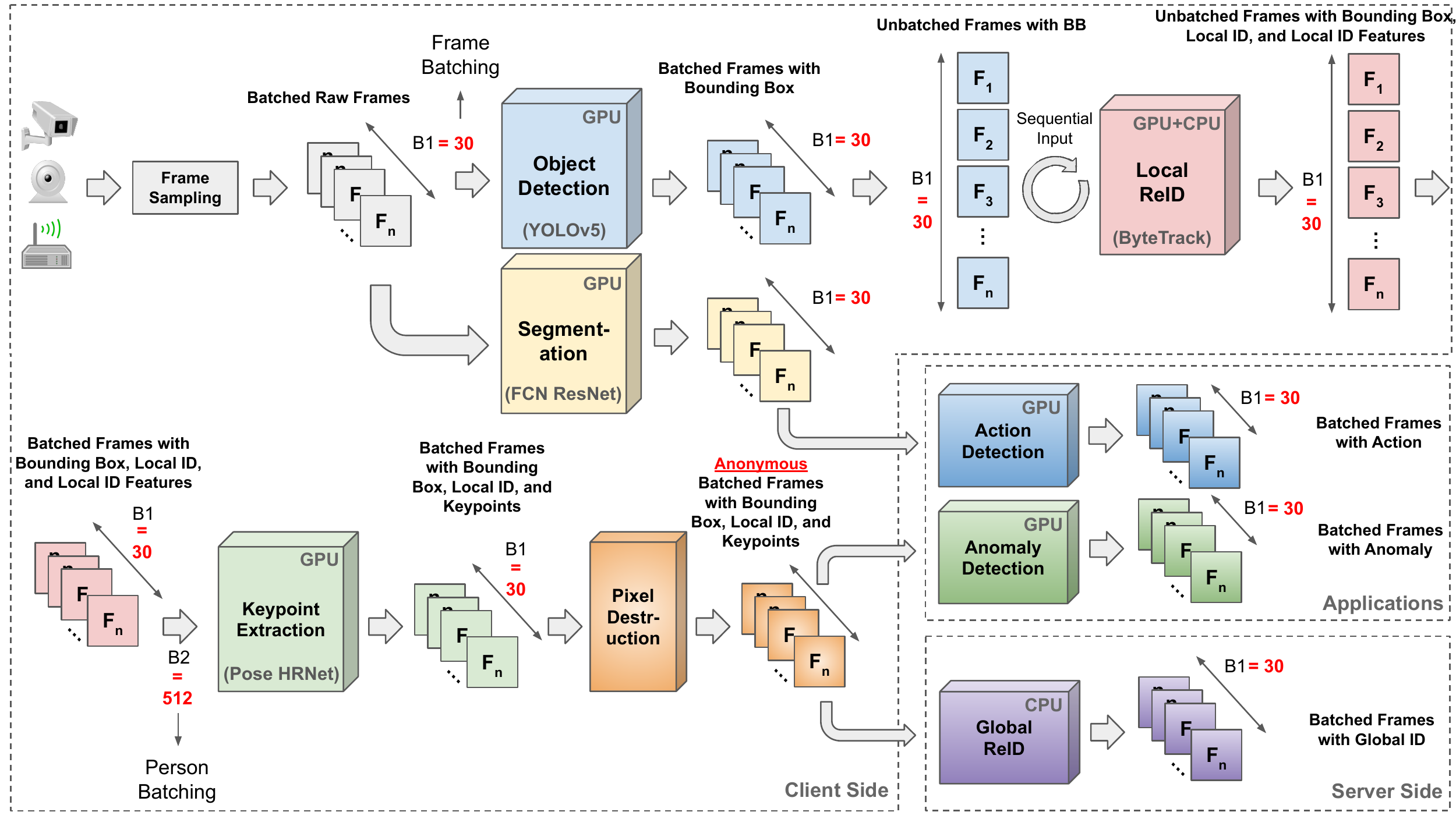}
                    \caption{Pipeline flow block diagram. \cite{glenn_jocher_2022_7347926, zhang2022bytetrack, sun2019deep, xiao2018simple, 7478072}}
                    
                \label{fig:pipe}
\end{figure*} 

As we can see in figure \ref{fig:pipe}, the pipeline of our system consists of different algorithms. The pipeline receives videos from cameras and uses YOLOv5 as the object detection service to detect the persons in each frame  \cite{glenn_jocher_2022_7347926}. The YOLOv5 service puts a bounding box around the detected person. The ByteTrack is used for assigning the IDs to each object by re-identifying them between frames. These IDs are crucial to re-identify the detected objects for each camera and enable the system to track objects detected by a camera \cite{zhang2022bytetrack}. This system cannot affect society's safety without algorithms that detect actions and anomalous behaviors. Pixel-based and pose-based algorithms are two main approaches to detect actions and anomalies \cite{8853267}. Pose-based algorithms are preferred for the objectives of this system since by using the pose-based algorithms, the face of people is not recognized, and the results are gender and racial-neutral. We use Pose HRNet to extract the person's keypoints and estimate the human pose \cite{sun2019deep, xiao2018simple}. We also use Graph Embedded Pose Clustering (GEPC) \cite{markovitz2020graph} for the anomaly detection task.
On the other hand, Real-World Graph Convolution Networks (RW-GCNs) \cite{DBLP:journals/corr/abs-2201-05739} are used for the action recognition task. Anomaly and action detection are just some of the tasks that use the keypoints. To achieve the goal of this system, we should be able to track people across different cameras. Global re-identification task uses the extracted keypoints within multiple cameras in a newly seen domain \cite{ye2021deep}. Online learning on the abstract feature representation is an essential part of re-identification across multiple cameras. Online learning also helps with privacy preservation as after a short amount of time, usually within minutes, a new model based on new data is trained, and all the old weights are destroyed, which mainly helps with keeping the stored data encoded.\cite{7090533}

The pipeline results will be stored in a local database hosted by the physical server. Global IDs (re-identified detection between multiple cameras), camera IDs, record time, bounding box information, anomaly scores, and the recognized actions associated with each global ID are stored in the local database. It is a very important feature of our system that we do not store images nor in the local database neither in any stages of the pipeline. We then analyze the data to extract the information required to deliver it to the end users. We use the global IDs, camera IDs, and record time features to calculate the real-time number of people at each camera, the total number of people tracked by each camera over time, and the cumulative number of people across all cameras installed in the location. We calculate the real-time occupancy at each camera and show it on an occupancy indicator. This feature shows the current number of people at each location compared to the historical appearance of people respecting the time. The bounding box information includes X and Y coordinates of the top left corner, width, and height of the box. We use the bounding box information to generate the real-time Bird's-Eye-View (BEV). BEV feature provides us the information on how people are using spaces at each location, and it also enables us to detect group movements as well as group behaviors. Then we merge the BEVs of the last 24 hours to generate the heat map in each location.
Moreover, another essential function of the system is to notify the users in an emergency. We define the emergency case as detecting specific objects such as guns, detecting anomalous behavior such as fighting, and observing an unexpected amount of people. As it is shown in Figure \ref{fig:view_points}, the analysis conducts on the local server. 

The results of the analysis are sent to a cloud server. Using a cloud-native service to implement the smartphone application provides robust data storage and management, scalability, and user management solutions\cite{dahunsi2021commercial}. In this real-time system, it is crucial to lowering the latency as much as possible. We use a low-latency database to store the data points on the cloud. The low-latency database enables the application developer to query within the stored data with the key-value attribute\cite{ani11092697}. Moreover, the users can serach easily in the database for the desired statistics over time and in different locations. We chose the camera ID and timestamp as the key-values of tables. Since there are several cameras in each location and we need to push real-time data to the cloud server, we need a gateway to ensure each data point is pushed correctly to the specified table of the database. This gateway generates topic rules based on the key-value to push data directly to the specified database. Another function of the gateway is to ensure the users are notified in emergency cases via their devices. Therefore, necessary topics and messages are created on the gateway to enable the service communicate with the rule-based message service. In the analysis section on the physical server, the emergency cases such as detecting anomalous behavior are distinguished and are pushed to the rule-based gateway as a specific topic. The gateway then communicates with rule-based message service to publish the message on users' devices. 
We also need an application development service that enables the application developer to generate required APIs for the smartphone application. To manage the users we use an authentication service on cloud.

\section{Evaluation}
In this section, we will elaborate on the evaluations of the proposed system. This system could be evaluated on two bases: quantitative and qualitative. In the quantitative subsection, we will report the results of the algorithms we utilized. In the qualitative subsection, we will argue how this early system can be considered a baseline model for improving public safety in SVS.

\subsection{Quantitative Evaluation}

In order to achieve the goal of the system, we need to use an appropriate object detection algorithm. In our case, we use YOLOv5 \cite{glenn_jocher_2022_7347926}. It accepts frames as input and will provide bounding boxes for the objects it sees, including the pedestrians in each frame. There are multiple versions of YOLOv5 available, but to achieve a fair balance between speed and accuracy, we use YOLOv5x, which has an mAP (val 50) of  $68.9\%$ for single-model-scale on COCO val2017 \cite{lin2014microsoft}. The speed on the Nvidia V100 GPU with a batch size of 32 is $4.8ms$ averaged over COCO val images.

Following the object detector, we use ByteTrack \cite{zhang2022bytetrack} for re-identifying people between consecutive frames. ByteTrack reports a MOTA of $80.3$ and $77.8$ on MOT17 and MOT20, respectively, from MOT challenge test set \cite{MOTChallenge2015, MOTChallenge20, MOT19_CVPR}. The reported throughput is $29.6$ and $13.7$ on MOT17 and MOT20, respectively.

In order to obtain the keypoints of each person, we use Pose HRNet \cite{sun2019deep, xiao2018simple}. Pose HRNet is a top-down approach for human pose estimation, receiving the bounding boxes from previous stages and outputting the 17 COCO style keypoints. We use HRNet-W48 which has an average precision of $76.3\%$ on the COCO validation set and an average precision of $77.0$ on the COCO test-dev set. 

These final keypoints can be utilized for any downstream tasks such as anomaly detection, action detection, and global re-identification. Global re-identification refers to re-identifying people between multiple cameras without revealing any biases toward identifying the characteristics and demographics of each person, such as race, ethnicity, and color.

As an example of downstream tasks, we use GEPC \cite{markovitz2020graph} for anomaly detection. GEPC utilizes the obtained keypoints in order to identify anomalous behaviours if they exist in a frame. GEPC provides two different approaches for anomaly detection namely GEPC-Pose and GEPC-Patches. GEPC-Patches uses pixel information and thus not useful in our case. GEPC-Pose on the other hand only uses keypoints information with and on the ShanghaiTech dataset \cite{DBLP:journals/corr/abs-1807-09959} reported Area Under the Receiver Operating Characteristic curve of 0.752.

Any other application that uses pose information can be another downstream application of the proposed system. 
Table \ref{tab:quant} represents the summary of the quantitative evaluations. 
\begin{table}[]
\centering
\caption{Summary of Quantitative Evaluation.}
\label{tab:quant}
\begin{tabular}{c|c|c|c}
\multicolumn{1}{c|}{\textbf{algorithm}} & \textbf{data set} & \textbf{metric} & \textbf{results} \\ \hline\hline
YOLOv5                                   & COCO             & accuracy        & 68.9\%           \\ \hline
ByteTrack                                & MOT17            & MOTA            & 80.3             \\ \hline
ByteTrack                                & MOT20            & MOTA            & 77.8             \\ \hline
ByteTrack                                & MOT17            & throughput      & 29.6             \\ \hline
ByteTrack                                & MOT20            & throughput      & 13.7             \\ \hline
HRNet-W48                                & COCO val         & precision       & 76.3             \\ \hline
HRNet-W48                                & COCO test-dev    & precision       & 77               \\ \hline
GEPC-Pose                                & ShanghaiTech     & AUROC           & 0.752           
\end{tabular}
\end{table}

\subsection{Qualitative Evaluation}
\begin{figure}[]
        \centering
               \includegraphics[width=1\linewidth]{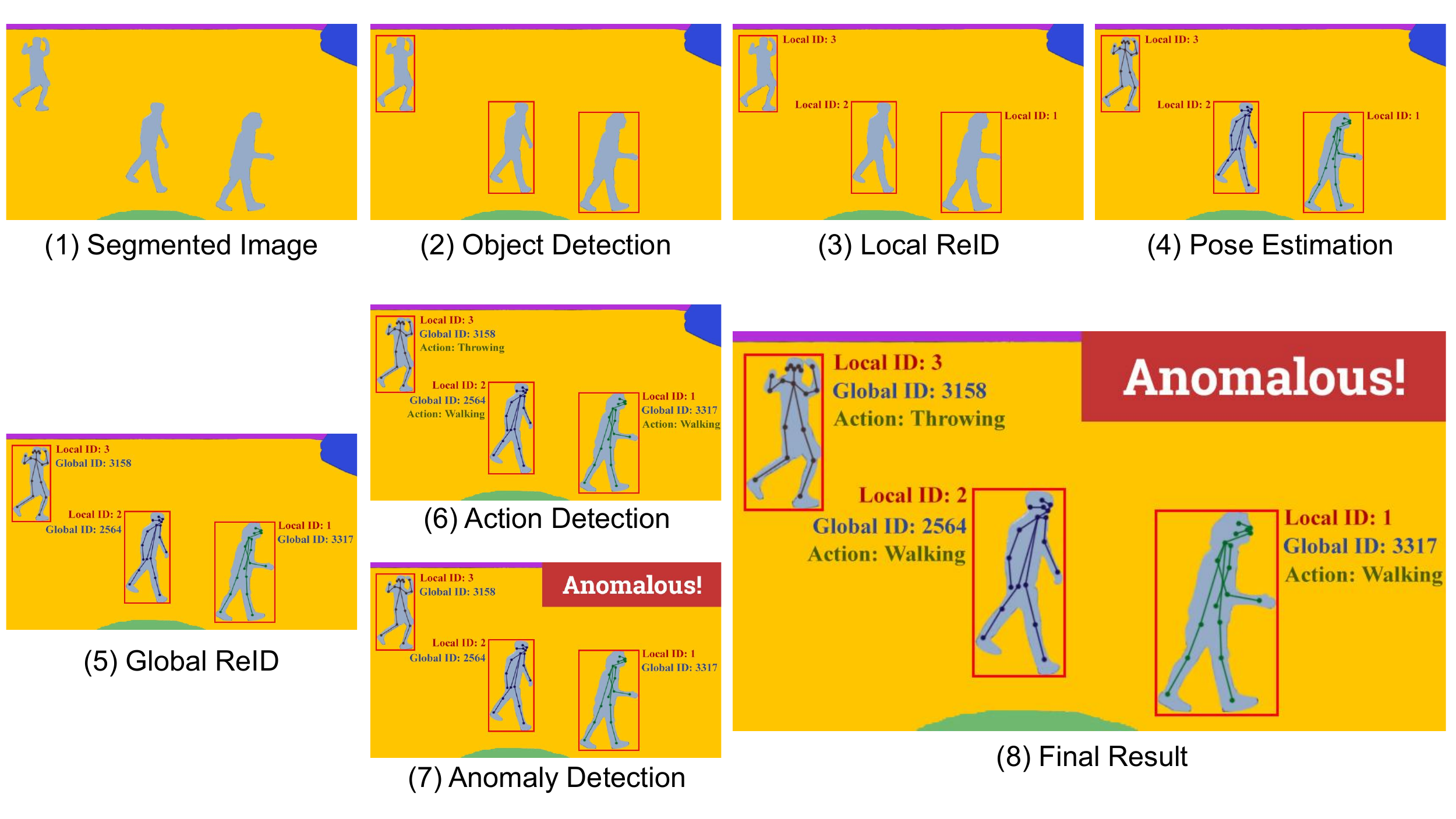}
                    \caption{Pipeline Qualitative Representation.}                    
                \label{fig:qual}
\end{figure} 
We discussed in the proposed system section the algorithms and services and the data flow of this design. We evaluate the system qualitatively based on the four levels of privacy perseverance discussed in the Privacy Perseverance System Features section. From the algorithm perspective, all AI-based algorithms are pose-based algorithms. In this design, we avoid using algorithms that use identifiable information, such as pixel-based algorithms. The most important parts of the system, such as anomaly detection, action recognition, and global re-identification, use skeleton and abstract feature representation. As a result of this approach toward the algorithm, neither the inputs nor the outputs are identifiable information. Moreover, the outputs are race, gender, age, and ethnicity neutral. This specific aspect of the system addresses the issue of discrimination as a fundamental ethical challenge in the public safety domain\cite{nissenbaum2004privacy}.

Data transmission is an essential aspect of designing systems in the SVS context \cite{nissenbaum2004privacy, hartzog2018privacyos}. Images of people are accessible through cameras in SVS systems. These images can be used directly by the data collector for processing purposes or can be sold by the data collector to a third party. Whether the data is used by the data collector or sold to a third party, the system's designer should consider the essential security practices to prevent image and data leaks. Although our system is not dependent on identifiable information, we should ensure that the information can not be transferred to an unauthenticated party. To address this issue, we are using a local server protected through different firewalls, and only verified users can access the server. As discussed earlier in this paper and Figure \ref{fig:view_points}, the SVS pipeline and de-identified information database are hosted on a local server. We also analyze the data on the local server and only push the analyzed data to a cloud-based server.

Data retention and irreversibility are the focus of all privacy protection acts. This means outputs should not be identifiable nor reversible when selecting the proper models for the machine learning tasks. We are following two approaches to consider this issue. First, we are not using facial recognition technologies. As described in the Proposed System section, we use pose-based models in all pipeline sections. Using pose-based models ensures that our system is not using identifiable information, i.e., images.
Moreover, we are not storing the image frames in any section of the system; therefore, no one has access to the images captured by the cameras. Although we are using abstract feature representations to train the models, there might be a problem of reversibility in the global re-identification model. The global re-identification model uses the features of the bounding boxes to identify objects across multiple cameras\cite{ye2021deep}. Therefore, the global re-ID model should be able to store these features and, once it detects a person cross check the features of the new person with the stored features to decide whether this one is a new person. From the model perspective, these features can be identifiable by reversing back. Theoretically, if someone has access to the abstract feature representations of each person and the weights of the neural network model, S/he can decode the model and recover the main image to an acceptable resolution level \cite{radenovic2018fine}. Our second approach is to resolve this reversibility problem. Our proposed solution is using online learning for global re-ID. This approach updates the model weights every 30 minutes, and the previous weights will be destroyed automatically. Therefore there is no chance to revert the images even if someone has access to the extracted features. 

According to privacy protection acts, the type of stored data is important for a system to be compliant with these acts. As we discussed earlier in this paper, We have not only remarked on considerations regarding the stored data but also have considered the ethical considerations at the data processing level. As described before, we are not using pixel-based algorithms according to our system features. Moreover, we need to store and transfer the actual videos or images. Putting these two together makes the output data de-identified. On the other hand, we are not using facial recognition technology in local and global re-ID algorithms. This guarantees that we do not use any personally identifiable information to detect persons. 

Figure \ref{fig:qual} represents the qualitative figure of pipeline.
Table \ref{tab:challenge} summarizes the features of our proposed system to address the ethical challenges of utilizing SVS to improve public safety. 

\begin{table}[]
\centering
\caption{The proposed system's solutions for policy challenges.}
\label{tab:challenge}
\begin{tabular}{c|c}
\textbf{Metrics}   & \textbf{Solution} \\ \hline \hline
Algorithm  & Using Pose-Based Algorithms  \\ \hline
System     & Using Local Server    \\ \hline
Model     & Making Data Irreversible     \\ \hline
Data    & Not Using PII      
\end{tabular}
\end{table}

\section{Conclusion and Discussion}
Monitoring the public's activities is a typical approach to improve public safety. However, monitoring public behaviors inherently brings an ethical challenge\cite{miller2017ethical}. There is always a trade-off between the limits of monitoring and respecting people's privacy\cite{crow2017community}. While someone believes that by escalating monitoring systems, everyone can benefit from a safer society, others argue that violating privacy at the expense of safety is not a desirable choice \cite{townsend2011privacy}.   

Incorporating AI in the form of SVS systems into the current passive surveillance systems exacerbated this ethical challenge. There are some regulations and policies that address this issue. However, these regulations provide a general and holistic overview of using these technology and are considered the baseline of the system design approach. Therefore, the concerns about privacy violations of these systems require special attention to the design of these systems. 
We proposed an early design of an end-to-end system that creates a road map for a more holistic perspective toward addressing ethical challenges in designing an SVS system. There are some specific privacy challenges in the SVS. Using PII and facial recognition technology in processing can violate people's privacy. Storing and transferring actual videos can enhance the possibility of privacy violations. 
We argued that privacy issues in designing an SVS system could be addressed through four perspectives: Algorithm, system, model, and data. From the algorithm perspective, to ensure that we are not using identifiable information, we utilize pose-based algorithms. In order to increase the security of the system, we run all pipeline algorithms on a local server. We also use online learning methods for local and global re-identification as a solution to reversible data. From a data perspective, we are not storing images and videos.

We argued that the system is generally set up to address the privacy challenges; However, privacy is only one of the ethical aspects the current system addresses. Discrimination is currently a critical ethical issue in policing society\cite{miller2017ethical}. Since we are not using PIIs and facial recognition technology, our system is racial, age, and gender neutral, which provides a fundamental baseline to remove biases in the policing and monitoring processes. 

As shown in the Evaluation section, our end-to-end system is fully functional. However, the results still need to be state-of-the-art and improved. Indeed, in this early stage, our focus is on two aspects: the functionality of the system and addressing the ethical issue of privacy. Therefore, In the current setup, we are dropping the accuracy to respect privacy. Optimization of the algorithms and services of this system on the local server and cloud server will be the next step. 

This system can be used in various public domains to help the community improve public safety. It also can be used by the public and private sectors as a more efficient alternative to the current passive surveillance systems. Public parking lots, grocery store parking lots, university campuses, bus and train stations, the downtown of the cities, and plazas are examples of the sectors that can use this system to improve their surveillance system. Although safety assurance through a privacy perseverance system is the system's primary goal, both private and public sectors can benefit from the information provided. That information can provide insightful business solutions for the sector. 

Improving the system's overall functionality, improving the accuracy, lowering the latency, and optimizing the bandwidth usage and CPU and GPU usage are possible future works from the system perspective. From the data perspective, more advanced statistical analysis is required. Considering the currently provided data, such as bird's eye view, heat map, action recognition, and anomaly detection, studying various social issues respecting the occupied spaces might be very interesting for sociologists. For example, if any individual or group actions are more likely to occur in a specific place. Studying the factors that can result in engaging communities with this system is another aspect that could be addressed. Finally, since we only focused on privacy issues as the ethical challenges in the context, there is still room to study the effect of other ethical issues, such as trust, in designing SVS systems for delivering safety to society.

\bibliography{aaai23}

\end{document}